\documentclass[a4paper]{article}

\usepackage{amsmath,amsthm,amsfonts,graphicx}
\usepackage{natbib}

\newtheorem{theorem}{Theorem}
\newtheorem{condition}{Condition}

\begin{document}

\begin{center}
{\Large\bf On a non-parametric confidence interval for the regression slope}\\
(Running title: \`A la Tukey confidence interval for slope)

\bigskip
\noindent
R\'obert T\'oth$^1$ and J\'an Somor\v{c}\'ik$^2$\long\def\symbolfootnote[#1]#2{\begingroup\def\thefootnote{\fnsymbol{footnote}}\footnote[#1]{#2}\endgroup}\symbolfootnote[1]{Corresponding author: somorcik@fmph.uniba.sk}\\
$^1$Tangent Works, Na Slav\'ine 1, 81104 Bratislava, Slovakia\\
$^2$Comenius University Bratislava, Mlynsk\'a dolina, 84248 Bratislava, Slovakia
\end{center}

\begin{abstract}
We investigate an application of the Tukey's methodology in Theil's regression to obtain a confidence interval for the true slope in the straight line regression model with not necessarily normal errors. This specific approach is implemented since 2005 in a package of the software R; however, without any theoretical background. We illustrate by Monte Carlo simulations, that this methodology, unlike the classical Theil's approach based on Kendall's tau, seriously deflates the true confidence level of the resulting interval. We provide also rigorous proofs in case of four data points (in general) and in case of five data points (under some additional conditions); together with a real life methods usage example in the latter case. Summing up, we demonstrate that one should never combine statistical methods without checking the assumptions of their usage and we also give a warning to the already wide  community of R users of Theil's regression from various fields of science.
\end{abstract}


\section{Introduction}

The Theil's regression (sometimes referred to as Theil--Sen regression) is a robust non-parametric replacement of the traditional least squares approach to the straight line regression model $Y=\beta_0 + \beta_1 x + \varepsilon$ and also to some more complex linear regression models (the pioneering papers were \citealt{theil1950_part_I,theil1950_part_II,theil1950_part_III}). The Theil's methodology does not require normality of the random errors $\varepsilon$, while being able to provide parameter estimates, tests of linear hypotheses about the parameters, as well as confidence intervals for the parameters (see e.g. \citealt{hollanderwolfe1999} for a detailed description of the methods).

We focus on the confidence interval (CI) for the true slope $\beta_1$. For the software R (\citealt{eRko}) there exists a package called mblm (\citealt{package_mblm}) that includes many tools of Theil's regression. But, surprisingly, when asked for a CI for $\beta_1$, the package does not compute the classical Theil's CI for $\beta_1$, proposed already in \citet{theil1950_part_I} and making use of the theory of Kendall's tau. Instead, the package uses a different approach that utilizes without any reference the well-known CI based on the Wilcoxon's signed rank test. In general setting, the CI based on the Wilcoxon's signed rank test has been ascribed to John Tukey (see \citealt{hollanderwolfe1999} for historical details), who originally developed it to obtain a CI for the true center of symmetry of a symmetric distribution from which we observed a sample of \emph{independent} and identically distributed data. However, it turns out very quickly (see Section~\ref{section_new_interval}), that in case of slope estimation in Theil's regression the input data are definitely \emph{not independent}. Therefore, the true confidence level of the resulting interval provided by the package mblm is of question and our paper shows that this negative premonition turns real. We think that it is important to point out and study this issue; and not just for theoretical reasons, i.e. to demonstrate that even a tempting combination of some proven statistical methods may have disastrous consequences if one ignores the assumptions of their usage. Our practical motivation was to alert the community of applied statisticians, because Theil's regression (for its simplicity) and the corresponding R package mblm (for its availability) are rather popular among researchers who apply statistics in other sciences -- see e.g. \citet{*17*logan2011}, a textbook for biologists that recommends the package mblm, or this sample of papers reporting the use of the package mblm in microbiology, genetics, chemistry, ecology, forestry, agriculture, hydrology, meteorology and also in behavior analysis:
% microbiology, genetics, chemistry:
\citet{*21*hunter_etal2012},      %mikrobiologia - vydavatel=nature.com
\citet{*23*carothers_etal2010},   %RNA
\citet{*06*kumari_etal2012},      %gene_association_PLOS_ONE
%
% ecology:
\citet{*24*denys_etal2012},       %arsen_cadmium
\citet{*15*hunter_etal2013},      %podmorsky_kanon
\citet{*19*lucas_etal2013},       %kationy_v_riekach
%
% forestry, agriculture:
\citet{*11*sardans_penuelas2015}, %rast_stromov
\citet{*09*pocewicz_etal2007},    %borovicovy_les
\citet{*07*heiskanen_etal2011,    %tajga
       *10*heiskanen_etal2012},   %tajga2
\citet{*04*mueller_etal2014},     %land_use
\citet{*14*arpaci_etal2013},      %fire_weather_indices
\citet{*05*eastaugh_etal2012},    %fire_danger_indices
\citet{*18*barroso_etall2015},    %brazilska_quantile_regression
%
% hydrology, meteorology:
\citet{*13*cuevas_etal2010},      %slovenska hydrologia
\citet{*03*zottele_etal2010},     %talianska_agrometeorologia
\citet{*16*puertas_etal2011},     %cilania_zrazky
%
% behavior analysis:
\citet{*08*vannest_etal2013}.    %kniha_Single Case Research in Schools

Our paper is organized as follows. Section~\ref{section_model} defines the underlying model. In Sections~\ref{section_theil} and \ref{section_new_interval} we provide detailed description of the classical Theil's CI and the CI based on the Tukey's methodology, respectively; with an illustrative application of both methods on a real life dataset of $n=5$ data points in Section~\ref{section_example}. Section~\ref{section_simulacie} shows by means of Monte Carlo simulations that the CI based on the Tukey's methodology has its true confidence level under the nominal confidence level which is set to the traditional $95\%$ throughout the whole paper. We prove this observation rigorously in case of $n=4$ data points (Section~\ref{section_n=4}). In Section~\ref{section_n=5}, under some additional conditions, we provide a proof also in the setting of the above-mentioned real life example, i.e. for $n=5$ data points. For the sake of completeness we treat also the case of $n=3$ data points in Section~\ref{section_n=3}. Finally, in Section~\ref{section_mblm} we add some notes on the R package mblm implementation of the CI for the true slope based on the Tukey's methodology. Proofs of the theorems were deferred to the Appendix.

%___________________
\section{The model}\label{section_model}

For each of the $n$ fixed and distinct points $x_1,x_2,\ldots,x_n$ (values of the predictor $x$) we observe the value of a random variable $Y$ (response). We get a set of observations $Y_1,Y_2,\dots,Y_n$, where $Y_i$ is the response at $x_i$. Without loss of generality we assume that $x_1<x_2<\cdots<x_n$. Our linear model has the form
\[
 Y_i=\beta_0+\beta_1 x_i+\varepsilon _i, \quad i=1,2,\ldots,n,
\]
where $\beta_0$ (intercept) and $\beta_1$ (slope) are unknown parameters. Finally, the unobservable random errors $\varepsilon_1,\varepsilon_2,\ldots,\varepsilon_n$ are iid random variables from a continuous (not necessarily normal) distribution.

%___________________
\section{Theil's confidence interval for slope}\label{section_theil}

The hypothesis $$H_0: \beta_1=\beta^*$$ can be tested using $D_i=Y_i-\beta^* x_i=(\beta_1-\beta^*)x_i+\beta_0+\varepsilon_i$. Provided that $H_0$ is true, the $D_i$'s do not depend on the $x_i$'s, i.e. they do not correlate. Hence, the validity of $H_0$ can be ``measured'' e.g. by the sample Kendall's correlation coefficient
\[
\tau = \frac{N_c-N_d}{\binom{n}{2}},
\]
where $N_c$ is the number of concordant pairs (i.e. pairs of points $[x_i,D_i]$ and $[x_j,D_j]$ such that $(x_i-x_j)(D_i-D_j)>0$) and $N_d$ is the number of discordant pairs (i.e. pairs of points $[x_i,D_i]$ and $[x_j,D_j]$ such that $(x_i-x_j)(D_i-D_j)<0$). The test statistic $K = \binom{n}{2} \tau = N_c-N_d$ is known as the Kendall $K$ statistic (see \citealt{hollanderwolfe1999}). By $\mathbb{K}_n$ we denote its distribution under independence of the $D_i$'s from the $x_i$'s. The distribution $\mathbb{K}_n$ has been tabulated (see e.g. \citealt{hollanderwolfe1999}) and implemented in many statistical softwares (see e.g. \citealt{package_Supp_Dists}), because it depends just on the sample size $n$, but not on the distribution of the data. The distribution $\mathbb{K}_n$ is discrete, symmetric and has the support
\[
 \left\{-\binom{n}{2},-\binom{n}{2}+2, -\binom{n}{2}+4 ,\ldots,\binom{n}{2}-4,\binom{n}{2}-2, \binom{n}{2}\right\},
\]
because $K$ has the same parity as $\binom{n}{2}$. A test of the hypothesis $H_0$ at the significance level $\alpha$ is then
\[
\text{reject } H_0:\beta_1=\beta^* \text{, if }|K| \geq k_n(\alpha/2),
\]
where $k_n(\alpha/2)$ stands for the upper quantile of the distribution $\mathbb{K}_n$. It should be such an integer that, under $H_0$, $P(K \geq k_n(\alpha/2)) = \alpha/2$. However, due to the discrete nature of the distribution $\mathbb{K}_n$, an exact equality is virtually impossible. Therefore, we define $k_n(\alpha/2)$ as such a unique integer with the same parity as $\binom{n}{2}$ that
\[
 P(K \geq k_n(\alpha/2)) \leq \alpha/2 \quad \text{and} \quad P(K \geq k_n(\alpha/2) - 2) > \alpha/2.
\]
The consequence is that in general the true probability of the type I error of the above test is bellow the nominal significance level $\alpha$, because it equals $2 \cdot P(K \geq k_n(\alpha/2))$.

Similar idea leads to a CI for the true slope $\beta_1$. Denote
\[
 S_{ij} = \frac{Y_i - Y_j}{x_i - x_j}, \quad (i<j)
\]
the ``sample'' slope of the line given by the pair of sample points $[x_i,Y_i]$ and $[x_j,Y_j]$. There are $N = \binom{n}{2} = n(n-1)/2$ such slopes. Order them ascendingly and denote the resulting sequence $s_1<s_2<\cdots<s_N$ -- since the random $\varepsilon_i$'s come from a continuous distribution, we may ignore ties between the sample slopes, because they will happen with zero probability, i.e. we shall assume that there are sharp inequalities between the $s_i$'s. The \emph{Theil's $1-\alpha$ confidence interval} for the true slope $\beta_1$ is
\begin{equation}\label{theil_interval}
 (s_l,s_u),
\end{equation}
where
\[
l=\frac{N-k_n(\alpha/2)}{2}+1
\quad \text{ and } \quad
u=\frac{N+k_n(\alpha/2)}{2};
\]
see e.g. \citet{hollanderwolfe1999}. Note that the indices $l$ and $u$ are symmetric in the sense that the same CI is obtained also by taking the $l$-th slope from bellow a the $l$-th slope from above, since $l+u=N+1$.
The discrete nature of the distribution $\mathbb{K}_n$ involved causes the true confidence level of the above interval to be typically over $1-\alpha$. The exact value is given by the following theorem.
\begin{theorem}\label{theorem_theil_confidence}
For all $k$ of the form $N-2i$ ($i=0,1,\ldots,\lfloor N/4 \rfloor$) put $l=(N-k)/2+1$ and $u=(N+k)/2$. Then the true confidence level of the Theil-type CI $(s_l,s_u)$ is $1 - 2 \cdot P(K \geq k)$ where $K$ is a random variable following $\mathbb{K}_n$.
\end{theorem}
Theorem~\ref{theorem_theil_confidence} implies that the true confidence level of the Theil's CI~(\ref{theil_interval}) equals
\begin{equation}\label{theil_confidence}
1 - 2 \cdot P(K \geq k_n(\alpha/2)),
\end{equation}
which is at least $1-\alpha$.

%___________________
\section{An \`a~la~Tukey confidence interval for slope}\label{section_new_interval}

We start with a brief description of the Tukey's methodology in a general setting (see e.g. \citealt{hollanderwolfe1999}).

Suppose we have some input iid random variables $Z_1,Z_2,\ldots,Z_N$ coming from a conti\-nuous symmetric distribution. We compute the so-called Walsh averages $(Z_i+Z_j)/2$ ($i \leq j$). Now we order the Walsh averages ascendingly (due to the continuity of the underlying distribution, we may ignore ties) and denote the resulting set $w_1 < w_2 < \cdots < w_P$, where $P=\binom{N}{2}+N = N(N+1)/2$.  Then the Tukey's $1-\alpha$ CI for the center of symmetry of the true distribution of the $Z_i$'s will be
\begin{equation}\label{tukey_interval}
 (w_{L},w_{U}),
\end{equation}
where $U=t_N(\alpha/2)$. The value of $L$ will be ``symmetric'' in the sense that $L=P-t_N(\alpha/2)+1$, i.e. one takes the $t_N(\alpha/2)$-th Walsh average from bellow and the $t_N(\alpha/2)$-th from above. Finally, $t_N(\alpha/2)$ denotes the $\alpha/2$ upper quantile of the null distribution of the Wilcoxon's signed rank statistic $T^{+}$ having the range $0,1,2,\ldots,P$ (see e.g. \citealt{hollanderwolfe1999} for details). Similarly as with the Theil's CI, the true confidence level of the Tukey's CI is typically strictly above $1-\alpha$.

Now, as the R package mblm does, we apply the Tukey's methodology described above to obtain a CI for the true slope $\beta_1$ in Theil's regression. The role of the $Z_i$'s will be played by the set of the slopes of all lines given by all pairs of the data points, i.e. by the set $\{ S_{ij};i<j \}$. We name the resulting interval, i.e. the Tukey's CI based on the slopes $S_{ij}$, the \emph{\`a~la~Tukey confidence interval}. From a particular point of view, it seems to be a good idea to apply the Tukey's approach on the sample slopes $S_{ij}$, because the Tukey's approach was bred for and is known to perform well in situations of symmetrically distributed data -- and it is easy to see that the distributions of our $S_{ij}$'s are indeed symmetric aroud $\beta_1$! Unfortunately, one of the basic assumptions of the Tukey's methodology is independence of the input data, i.e. independence of the $Z_i$'s. However, it is easily seen, that this assumption does not hold for the slopes $S_{ij}$. Actually, there is  functional dependence among the slopes, because, for example, the knowledge of $S_{1,1},S_{1,2},\ldots,S_{1,n-1}$ enables us to compute the remaining $S_{ij}$'s, since
\[
 S_{ij}=\frac{S_{1j}(x_1-x_j)-S_{1i}(x_1-x_i)}{x_i-x_j}.
\]
This does not necessarily mean that the \`a~la~Tukey CI does not provide at least the nominal confidence level. For example, it may happen that the nominal confidence level is preserved, just the interval is redundantly wide. The best scenario from the \`a~la~Tukey CI's point of view is that the interval provides the nominal level of confidence while being narrower than the classical Theil's CI described in Section~\ref{section_theil}. The real state of affairs will be presented after an illustrative example.

%___________________
\section{A real life example}\label{section_example}

\begin{figure}[h]\centering
\caption{The cloud seeding experiment --- linear dependence of the double ratio $Y_i$ on time}
\label{figure_cloud_seeding}
\includegraphics{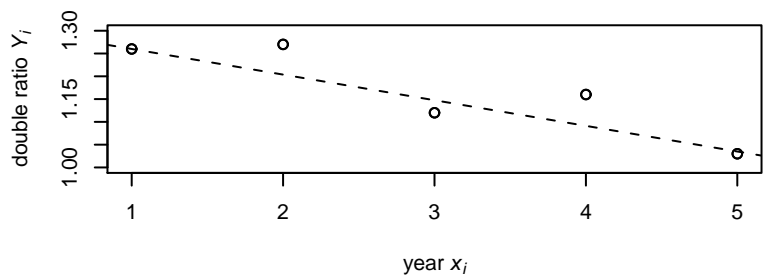}
\end{figure}

Let us explain the above-described methods on real data analyzed in \citet{hollanderwolfe1999} (Examples 9.1--3) and also in \citet{*17*logan2011} (Example 8.G), where the following description can be found. \citet{smith1967} investigated the effects of cloud seeding on rainfall in the Snowy Mountains, Australia. The experiment took place in two areas --- the target and the control. Within a year a number of periods were randomly allocated for seeding the target area and additional periods for non-seeding the target area. The total rainfalls $T_\text{seeded}, T_\text{unseeded}$ in the target and $C_\text{seeded}, C_\text{unseeded}$ in the control area during the seeding and non-seeding periods were recorded. Within a single year, the impact of seeding was assessed via a double ratio $Y=(T_\text{seeded}/C_\text{seeded}) / (T_\text{unseeded}/C_\text{unseeded})$ and the experiment was repeated over $n=5$ years (years denoted by $x$). The measurements are summarized in Table \ref{table_cloud_seeding} and depicted on Fig. \ref{figure_cloud_seeding}.

\begin{table}[h]\centering
\caption{The cloud seeding experiment --- the double ratio $Y_i$ measures the impact of cloud seeding on the rainfall in year $x_i$}
\label{table_cloud_seeding}
\begin{tabular}{c|cccccc}
year $x_i$ & 1 & 2 & 3 & 4 & 5 \\
\hline\noalign{\smallskip}
double ratio $Y_i$ & 1.26 & 1.27 & 1.12 & 1.16 & 1.03 \\
\end{tabular}
\end{table}

We adopt the classical straight line regression model $Y_i=\beta_0+\beta_1 x_i+\epsilon_i$ ($i=1,2,\ldots,5$). \citet{*17*logan2011} states that ``whilst there may not appear to be any evidence of non-normality\ldots, it could be argued that there are too few observations on which to make meaningful decisions about normality (of the random errors) and it might be safer to not make distributional assumptions''. Therefore, the Theil's regression in place of the classical least squares inference is applied. The ordered values of the $N=\binom{5}{2}=10$ sample slopes $S_{ij} = (Y_i - Y_j)/(x_i - x_j)$ are $s_1<s_2< \cdots < s_{10}$: $-.1500$, $-.1300$, $-.0800$, $-.0700$, $-.0575$, $-.0550$, $-.0450$, $-.0333$, $.0100$, $.0400$. The dashed line on Fig. \ref{figure_cloud_seeding} shows the linear trend estimated by the Theil's approach. Its slope is the median of the $S_{ij}$'s, i.e. $(-0.0575+(-0.0550))/2=-0.05625$ and suggests a decreas of the double ratio over time, i.e. a decreas over time of the rainfall increases resulting from the seeding.

Rather than a point estimate, our main concern are the $95\%$ CI's for the true slope. Let us start with the Theil's CI (Section~\ref{section_theil}). For $\alpha=5\%$ the appropriate upper quantile $k_5(2.5\%)$ is $10$ (see e.g. Table~A.30 in \citealt{hollanderwolfe1999}), $l=(10-10)/2+1=1$ and $u=(10+10)/2=10$. Hence, the resulting CI~(\ref{theil_interval}) is $(s_1,s_{10})$, i.e. it is given by the minimum and the maximum sample slope. Numerically,
\[
(s_1,s_{10})
=
(-0.15,0.04)
\]
and since the CI contains zero, the negative trend suggested by the point estimate of the true slope does not seem to be significant.

For the the \`a~la~Tukey CI, the appropriate upper quantile $t_{10}(2.5\%)$ is $47$ (see e.g. Table~A.4 in \citealt{hollanderwolfe1999}). The ordered values of the $P=10\cdot(10+1)/2 = 55$ Walsh averages $(s_i+s_j)/2$ ($i \leq j$) are $w_1 < w_2 < \cdots < w_{55}$:  $-.150, -.140,\ldots, .040$. Further, $U=47$ and $L=55-47+1=9$. Therefore, the resulting CI~(\ref{tukey_interval}) is $(w_9,w_{47})$, i.e. it is given by the $9$-th Walsh average from bellow and the $9$-th Walsh average from above -- which is the $47$-th from bellow, since there are together $55$ Walsh averages. For this dataset,
\[
(w_9,w_{47})
=
(-0.100,-0.015)
\]
and since this CI does not contain zero, it confirms the negative trend suggested by the point estimate of the true slope, i.e. a decrease over time of the rainfall increases resulting from the seeding. Note that, from this point of view, there is a discordance between the Theil's CI and the \`a~la~Tukey CI.

The \`a~la~Tukey CI $(-0.100,-0.015)$ is reported also in \citet{*17*logan2011}, because the book utilized the R package mblm. For this data, the \`a~la~Tukey CI is much narrower then the rather conservative Theil's CI $(-0.15,0.04)$. This is not just coincidence and in general the \`a~la~Tukey CI is not to be trusted because in what follows we show that there is a confidence level issue with it.

%___________________
\section{Monte Carlo study}\label{section_simulacie}

At the end of Section~\ref{section_new_interval}, a positive scenario was hypothesized, that the \`a~la~Tukey CI could be narrower than the classical Theil's CI. This was supported also by the real life example about cloud seeding in Section~\ref{section_example}. However, this benefit turns out to be worthless because there is a crucial problem with the \`a~la~Tukey CI's true confidence level, as can be seen from the results of a Monte Carlo study we have conducted. In Table~\ref{table_simulacie} (except the last column; see below) we provide simulation estimates of the true confidence levels of the \`a~la~Tukey CI for the true slope under various settings. The number of data points $n$ changed from $6$ to $200$. The true values of intercept $\beta_0$ and slope $\beta_1$ were set to $0$ and $1$, respectively. The iid random errors $\varepsilon_i$ were generated from the normal distribution $N(0,0.01)$, the Cauchy distribution with location parameter $0$ and scale parameter $0.1$, or the uniform distribution on the interval $(-0.2,0.2)$. The motivation for the scale parameters of the distributions was to make the spread of the $\varepsilon_i$'s comparable with the spread of the $x_i$'s, i.e. to achieve that the data points $[x_i,Y_i]$ do not produce an ideally straight line, nor resemble a shapeless data cloud. In the part ``Evenly spaced $x_i$'s'', the $x_i$'s created an equidistant design on the interval $(0,1)$, more precisely, $x_i=(i-1)/(n-1)$ for $i=1,2,\ldots,n$. In the part ``Two clusters of evenly spaced $x_i$'s'', the design of the experiment consisted of two clusters of evenly spaced points on the subintervals $(0,1/3)$ and $(2/3,1)$, more precisely, $x_i=(i-1)/(3(n/2-1))$ for $i=1,2,\ldots,n/2$ and $x_i=2/3+(i-n/2-1)/(3(n/2-1))$ for $i=n/2+1,n/2+2,\ldots,n$. Each figure in Table~\ref{table_simulacie} (except the last column) is based on $10{,}000$ simulations and it is the proportion of times (rounded to three decimal places) the \`a~la~Tukey CI covered the true slope $\beta_1$. The nominal confidence level was set to $95\%$.

Just for illustration, the rightmost column of Table~\ref{table_simulacie} contains the true confidence levels of the Theil's CI: these figures are not based on simulations, they have been computed by (\ref{theil_confidence}) using the R package SuppDists (\citealt{package_Supp_Dists}). Thanks to the distribution-free property of the Theil's CI, these true confidence levels depend just on the number of data $n$ (i.e. not on the design of the $x_i$'s, or on the underlying distribution of the random errors $\varepsilon_i$) and are, of course, always at least as high as $95\%$.

\begin{table}
\caption{True confidence levels of the $95\%$ Theil's CI (computed numerically) and the $95\%$ \`a~la~Tukey CI (simulation estimates under various arrangements of the $x_i$'s and various distributions of random errors).\label{table_simulacie}}
\begin{tabular}{c|ccc|ccc|c}
\hline\noalign{\smallskip}
& \multicolumn{6}{c|}{\`a~la~Tukey CI} & Theil's CI\\
\noalign{\smallskip}\cline{2-7}\noalign{\smallskip}
& \multicolumn{3}{c|}{Evenly spaced $x_i$'s} & \multicolumn{3}{c|}{Two clusters of evenly spaced $x_i$'s}\\
\noalign{\smallskip}\cline{2-7}\noalign{\smallskip}
& \multicolumn{3}{c|}{Distribution of random errors:} & \multicolumn{3}{c|}{Distribution of random errors:}\\
Number of data $n$ & normal & Cauchy & uniform & normal & Cauchy & uniform \\
\noalign{\smallskip}\hline\noalign{\smallskip}
  6 & .869 & .850 & .867 & .871 & .855 & .867 & .983\\
 10 & .804 & .773 & .793 & .804 & .777 & .800 & .953\\
 20 & .679 & .636 & .675 & .678 & .646 & .675 & .953\\
 30 & .591 & .551 & .588 & .595 & .561 & .596 & .951\\
 40 & .533 & .494 & .540 & .530 & .499 & .541 & .952\\
 50 & .486 & .453 & .489 & .491 & .456 & .498 & .950\\
 60 & .449 & .416 & .451 & .452 & .424 & .457 & .950\\
 70 & .425 & .388 & .427 & .426 & .396 & .429 & .951\\
 80 & .399 & .365 & .402 & .402 & .373 & .403 & .951\\
 90 & .379 & .351 & .376 & .384 & .361 & .383 & .950\\
100 & .357 & .340 & .364 & .369 & .343 & .365 & .950\\
120 & .329 & .307 & .329 & .338 & .310 & .334 & .950\\
140 & .311 & .290 & .310 & .302 & .293 & .309 & .950\\
160 & .294 & .267 & .295 & .303 & .278 & .296 & .950\\
180 & .276 & .246 & .278 & .276 & .251 & .281 & .950\\
200 & .261 & .240 & .265 & .268 & .249 & .266 & .950\\
\end{tabular}
\end{table}

The main message of Table~\ref{table_simulacie} is that, irrespective of the probability distribution of the random errors, the true confidence level of the \`a~la~Tukey CI is strictly below the nominal $95\%$ and decreases rapidly with increasing number of data $n$ in all settings presented in our study. The design of the $x_i$'s does not seem to play an important role either; we tried also some other designs not reported here and the resulting figures were very similar. The Monte Carlo simulations support our suspicion that the method of construction of the \`a~la~Tukey CI for the true slope is wrong. In what follows we provide rigorous treatment of the problem in case of $n=5$, $n=4$, and $n=3$ data points.

%___________________
\section{The case of $n=5$ data points}\label{section_n=5}

We are going to examine the true confidence level of the \`a~la~Tukey CI for sample size and nominal confidence level as in the real life example about cloud seeding in Section~\ref{section_example}, i.e. in case of $n=5$ data points and nominal confidence level $1-\alpha = 95\%$. In Section~\ref{section_example} it was derived in detail, that the corresponding Theil's CI~(\ref{theil_interval}) and the \`a~la~Tukey CI~(\ref{tukey_interval}) are $(s_1,s_{10})$ and $(w_9,w_{47})$, respectively. The following theorem shows their mutual relationship.
\begin{theorem}\label{theorem_o_vnoreni}
For $n=5$, the $95\%$ \`a~la~Tukey CI $(w_9,w_{47})$ is always a subset of the $95\%$ Theil's CI $(s_1,s_{10})$.
\end{theorem}

Theorem~\ref{theorem_o_vnoreni} itself is not enough to claim that the \`a~la~Tukey CI has its true confidence level under $95\%$.  However, by Monte Carlo simulations not reported here we noticed that the \`a~la~Tukey CI $(w_9,w_{47})$ happens to be very often the subset of the even narrower Theil-type CI $(s_2,s_9)$. The true confidence level of $(s_2,s_9)$ can be obtained easily: put $l=2$, $u=9$, $n=5$, and $N=10$, then the notation of Theorem~\ref{theorem_theil_confidence} implies that $k=8$ and the theorem itself gives the true confidence $1-2 \cdot P(K \geq 8)$ which can be evaluated e.g. by Table~A.30 in \citet{hollanderwolfe1999}. The approximate result is $91.67\%$. Therefore, our aim is to show, that the \`a~la~Tukey CI $(w_9,w_{47})$ is ``very often'' a subset of $(s_2,s_9)$ with the poor confidence level of $91.67\%$. The consequence will be, that although the true confidence level of the \`a~la~Tukey CI $(w_9,w_{47})$ could be over that of $(s_2,s_9)$, it is definitely under $95\%$. The next theorem states exact conditions in terms of the $s_i$'s when the above-described desired ``very often'' inclusion of $(w_9,w_{47})$ in $(s_2,s_{9})$ happens, i.e. conditions when the lower (upper) bound of $(w_9,w_{47})$ is over (under) the lower (upper) bound of $(s_2,s_{9})$.

\begin{theorem}\label{theorem_o_t2_w9_w47_s_9}
If $n=5$ then: a) The random event $s_2 \leq w_9$ occurs if and only if $2s_2 \leq s_1+s_9$.
b) The random event $w_{47} \leq s_9$ occurs if and only if $s_2 + s_{10} \leq 2 s_9$.
\end{theorem}

The following theorem provides an upper bound for the true  confidence level $P(\beta_{1}\in (w_9,w_{47}))$ of the \`a~la~Tukey CI $(w_9,w_{47})$.
\begin{theorem}\label{theorem_o_p1+p2+p3+p4}
Let $n=5$. Denote
\begin{eqnarray}
p_1 & = & P(\beta_{1}\in (s_2,s_9)    \wedge 2s_2 \leq s_1 + s_9 \wedge s_2+s_{10} \leq 2s_9 ) \nonumber\\
p_2 & = & P(\beta_{1}\in (s_2,s_{10}) \wedge 2s_2 \leq s_1 + s_9 \wedge 2s_9 < s_2+s_{10}    ) \nonumber\\
p_3 & = & P(\beta_{1}\in (s_1,s_{9})  \wedge s_1+s_9 < 2s_2      \wedge s_2+s_{10} \leq 2s_9 ) \nonumber\\
p_4 & = & P(\beta_{1}\in (s_1,s_{10}) \wedge s_1+s_9 < 2s_2      \wedge 2s_9 < s_2+s_{10}    ) \nonumber
\end{eqnarray}
Then $P(\beta_{1}\in (w_9,w_{47})) \leq p_1+p_2+p_3+p_4$.
\end{theorem}

Let us discuss Theorem~\ref{theorem_o_p1+p2+p3+p4} in greater detail. The probability $p_1$ is simply the true confidence level of $(s_2,s_{9})$ together with the probability of the above-discussed ``very often'' inclusion of $(w_9,w_{47})$ in $(s_2,s_{9})$ with the poor true confidence level $91.67\%$. This means that $p_1$ is at most $91.67\%$. The probabilities $p_2$, $p_3$, $p_4$ are based on the true confidence levels of the slightly wider intervals $(s_2,s_{10})$, $(s_1,s_9)$, $(s_1,s_{10})$ and these confidence levels can be high. However, by Theorem~\ref{theorem_o_t2_w9_w47_s_9} it can be seen immediately that the probabilities $p_2$, $p_3$, and $p_4$ reflect only cases when these intervals include also $(w_9,w_{47})$ whereas $(w_9,w_{47})$ is not completely included in $(s_2,s_{9})$. Monte Carlo simulations suggest that these cases are rare, therefore we hope that $p_2$, $p_3$, and $p_4$ turn out to be low.

All we need to do is to determine the probabilities $p_1$, $p_2$, $p_3$, and $p_4$. Note that they are given just by the joint probability distribution of the ordered slopes $s_i$, i.e. one does not need to examine the much more complicated joint probability distribution of the ordered Walsh averages $w_i$ anymore. There are $10!$ possible orderings of the ten slopes $S_{1,2}, S_{1,3},\ldots,S_{4,5}$. Denote $B_1, B_2, \ldots,B_{10!}$ the appropriate random events, i.e. each $B_i$ denotes an event that a particular ordering of the slopes happened. Then, $p_1$ (and similarly $p_2$, $p_3$, and $p_4$) can be decomposed as
\[
p_1
=
\sum_{i=1}^{10!}
P(
 \{
  \beta_{1}\in (s_2,s_9)\wedge 2s_2\le s_1+s_9\wedge s_2+s_{10} \leq 2s_9 
 \}
 \cap
 B_i
 ).
\]
Let $B_i$ denote, for example, the ordering
\begin{equation}\label{konkretne_usporiadanie}
 S_{4,5} < S_{3,5} < S_{2,5} < S_{1,5} < S_{2,3} < S_{2,4} < S_{1,3} < S_{1,4} < S_{3,4} < S_{1,2}.
\end{equation}
Since
\begin{equation}\label{S_pomocou_epsilonov}
 S_{ij} = \frac{Y_i-Y_j}{x_i-x_j} = \beta_1 + \frac{\varepsilon_i-\varepsilon_j}{x_i-x_j},
\end{equation}
the $9$ inequalities in (\ref{konkretne_usporiadanie}) that define the ordering can be rewritten as $9$ linear inequalities of the form
\begin{equation}\label{linearna_nerovnoss_s_epsilonmi}
 c_1 \varepsilon_1 + c_2 \varepsilon_2 + \cdots + c_5 \varepsilon_5 < c_0,
\end{equation}
where $c_0,c_1,c_2,\ldots,c_5$ are constants depending on the $x_i$'s. Further, under $B_i$ the conditions $\beta_{1}\in (s_2,s_9)$, $2s_2\le s_1+s_9$ and $2s_9\ge s_2+s_{10}$ can be also rewritten as 2+1+1=4 inequalities of the form (\ref{linearna_nerovnoss_s_epsilonmi}), because under $B_i$, for example, $s_1=\beta_1 + \frac{\varepsilon_4-\varepsilon_5}{x_4-x_5}$, $s_2=\beta_1 + \frac{\varepsilon_3-\varepsilon_5}{x_3-x_5}$, etc. Therefore,
\[
p_1
=
\sum_{i=1}^{10!}
P(P_i),
\]
where $P(P_i)$ is the probability that the random errors vector $(\varepsilon_1,\varepsilon_2,\ldots,\varepsilon_5)^\top$ appears in the 5-dimensional polytope $P_i$ with faces given by the above-mentioned 9+4=13 linear inequalities of the form (\ref{linearna_nerovnoss_s_epsilonmi}).

Nevertheless, we still have to consider $10!=3{,}628{,}800$ polytopes and evaluation of $p_2$, $p_3$ and $p_4$ is going to quadruple this number. Fortunately, most of these polytopes are empty sets because the following theorem implies that a lot of the $10!$ possible orderings of the slopes are impossible.
\begin{theorem}\label{veta_obmedzujuca_usporiadania_sklonov}
Let $a<b<c$ be three indices. Then the slope $S_{ac}$ is neither the greatest nor the smallest of the trio $S_{ab}$, $S_{ac}$, and $S_{bc}$.
\end{theorem}
\noindent An automatized computer inspection of the $10!$ possible orderings revealed quickly, that only $768$ of them conform to Theorem~\ref{veta_obmedzujuca_usporiadania_sklonov}, which means that we have to deal just with $768$ polytopes to obtain a $p_i$. At this point we have to set concrete values of the $x_i$'s, because the $P(P_i)$'s depend on them. We decided for the following.
\begin{condition}\label{condition_ekvidistantnost}
The $x_i$'s create an equidistant design $x_i=i$ for $i=1,2,\ldots,n$.
\end{condition}
\noindent As byproducts, this choice of the $x_i$'s has the following pleasant consequences that again reduce the amount of computations.
\begin{theorem}\label{veta_nulovost_p3}
If Condition~\ref{condition_ekvidistantnost} holds and $n=5$, then $p_4=0$.
\end{theorem}
\begin{theorem}\label{p2=p3}
If Conditions~\ref{condition_ekvidistantnost} holds and the distribution of the random errors is symmetric (the latter and fairly common assumption will be posed later), then $p_2=p_3$.
\end{theorem}

Still, it is not easy to evaluate the $P(P_i)$'s under an arbitrary probability distribution of the random errors $\varepsilon_i$. Therefore, we decided for the uniform distribution to make the evaluation easier:
\begin{condition}\label{condition_rovnomerne_rozdelenie}
The probability distribution of the errors $\varepsilon_i$ is uniform on the interval $(-1,1)$.
\end{condition}
\noindent Under Condition~\ref{condition_rovnomerne_rozdelenie} the probability distribution of the vector $(\varepsilon_1,\varepsilon_2,\ldots,\varepsilon_5)^\top$ is uniform in the 5-dimen\-sional cube with the vertices $[\pm 1, \pm 1, \pm 1, \pm 1, \pm 1]$ and edges of length $2$. Therefore, the probabilities $P(P_i)$ reduce to
\[
 P(P_i) = \frac{V(Q_i)}{2^5},
\]
where $V(\cdot)$ denotes volume and the $Q_i$'s are the intersections of the polytopes $P_i$ (each given by a set of above-mentioned $13$ linear inequalities) with the cube given by the $10$ inequalities
\[
\varepsilon_i < 1 \text{ and } \varepsilon_i > -1 \quad (i=1,2,\ldots,5),
\]
which means that each $Q_i$ is again a polypote; given by $13+10=23$ inequalities of the form (\ref{linearna_nerovnoss_s_epsilonmi}). To evaluate the volumes of such polytopes we used a specialized software Vinci (\citealt{vinci}): for each polytope the $23$ defining inequalities were passed to Vinci and the computation of the volume was then based on the triangulation of the polytope and computation of determinants.

So we obtained $p_1$ and $p_2$ by evaluating the volumes of the above-mentioned $768+768$ polytopes and by Theorems~\ref{veta_nulovost_p3} and \ref{p2=p3} we have
\[
p_1 + p_2 + p_3 + p_4 = 0.8107315 + 0.0595787 + 0.0595787 + 0= 92{,}98889\%.
\]
This provides an upper bound for the true confidence level of the interval $(w_9,w_{47})$, i.e. the true confidence level is under the nominal level $95\%$ (as a Monte Carlo estimate based on $1{,}000{,}000$ simulations we obtained $87.9\%$). We note that Condition~\ref{condition_ekvidistantnost} is just technical, since we are able to evaluate $p_1+p_2+p_3+p_4$ under any particular arrangement of the $x_i$'s. However, Condition~\ref{condition_rovnomerne_rozdelenie} about the uniform distribution of the random errors is crucial, because it reduced our computation to evaluation of volumes of polytopes, which could be accomplished by Vinci.

%___________________
\section{The case of $n=4$ data points}\label{section_n=4}

In case of $n=4$ data points, the $95\%$ \`a~la~Tukey CI will be $(w_1,w_{21})$  which is obviously the same as $(s_1,s_6)$. Its true confidence level can be evaluated by Theorem~\ref{theorem_theil_confidence}: put $l=1$, $u=6$, $n=4$, $N=6$, obtain $k=6$ and the theorem gives the confidence $1 - 2 \cdot P(K \geq 6)$, which can be evaluated e.g. by Table~A.30 in \citet{hollanderwolfe1999}. The approximate result is $91.67\%$ which is definitely under $95\%$, i.e. the \`a~la~Tukey CI does not work correctly in this case either. Note that -- unlike the case of $n=5$ data points -- the obtained result $91.67\%$ holds in general, e.g. it is completely independent of the additional Conditions~\ref{condition_ekvidistantnost} or \ref{condition_rovnomerne_rozdelenie}.

The above paragraph also means that in case of $n=4$ data points the Theil's approach is unable to produce a $95\%$ CI, because the confidence level of $(s_1,s_6)$ (the widest Theil-type interval) is under $95\%$. From another point of view, the $95\%$ Theil's CI cannot be produced, because $k_4(2.5\%)$ satisfying our definition of the upper quantile value does not exist.

%___________________
\section{The case of $n=3$ data points}\label{section_n=3}

With just $n=3$ data points at hand, the Theil's approach breaks down, because $k_3(2.5\%)$ does not exist. The same happens to the $95\%$ \`a~la~Tukey CI, because the Tukey's methodology does not work for such a low number of data and nominal confidence level of $95\%$ ($t_3(2.5\%)$ does not exist).

%___________________
\section{An R implementation of the \`a~la~Tukey confidence interval}\label{section_mblm}

As we already noted, the \`a~la~Tukey CI for the true slope is implemented in the R package mblm, however, without any reference to a theoretical background. It is available in the CRAN package repository since 2005, but since that time the package documentation has been just noting that the package does not implement the original Theil's CI based on Kendall's tau and it is considered to be implemented in next version of the package. However, it has not been implemented till now  (August 2016), despite the fact that already the third version of the package has been released. Nevertheless, the main problem is that the package does not provide any warning about the deflated true confidence level of the intervals produced. The only exceptions are the cases of $n=4$ and $n=3$ data points. In case of $n=4$ data points the package mblm produces a correct warning message, that the requested confidence level is not achievable. However, careful inspection of the package code reveals that it is just a coincidence: in fact, the warning says nothing about the CI for the true \emph{slope}, because it has been invoked by the computation of a CI for the true \emph{intercept} (this CI was not discussed in our paper). For $n=3$ data points the package mblm produces an error message, however, as before the true reason for the message is a problem with the computation of a CI for the true intercept.

%___________________
\section{Conclusions}

We have shown by means of Monte Carlo simulations that the \`a~la~Tukey confidence interval for the true slope in the straight line regression model seems to be unable to achieve the nominal confidence level. The loss of interval's confidence does not seem to depend too much on the design of the experiment or on the distribution of the random errors, but becomes very serious with increasing number of data -- in all cases with over $160$ data points we observed the true confidence level even under $30\%$ instead of the nominal $95\%$.

In case of $n=4$ data points we easily obtained also the true confidence level of the \`a~la~Tukey confidence interval -- the simplicity of the reasoning resulted from the fact that the lower and upper limit of the \`a~la~Tukey confidence interval turned out to be some of the original sample slopes. However, in case of $n=5$ data points the situation was much more complicated: we were able to obtain only an upper bound for the true confidence level and we numerically evaluated this upper bound under the condition of uniformly distributed random errors.

Theoretically, the process of evaluation of the above mentioned upper bound can be adopted to obtain the exact value of the true confidence level of the \`a~la~Tukey interval. However, already in case of $n=5$ data points there are $55$ Walsh averages given by the ten slopes $S_{ij}$ and, theoretically, these Walsh averages can be arranged in $55!$ permutations. These would result in the necessity to evaluate and sum volumes of as much as $55! \approx 1.27 \cdot 10^{73}$ polytopes -- a very hard task from the numerical point of view. Similarly as in the evaluation of the $p_i$'s, many of these polytopes could be a priory shown to be of zero volume, but we decided to proceed in a different way: we estimated the true confidence level from above by terms not involving the Walsh averages and showed rather easily that this upper bound is strictly under $95\%$.

A natural question arises, if the reasoning in case of $n=5$ data points can be easily adopted or even generalized for larger $n$. Despite our effort we have not found any positive answer, because the situation complicates dramatically already for $n=6$.

The \`a~la~Tukey confidence interval for the true slope is implemented in the R package mblm without any warning about its deflated true confidence level. The results of our paper show that this functionality of the package (i.e. computation of the confidence interval for the true slope) should not be used, because it tends to provide too liberal interval estimates. We conclude that although the software R is of great help at a great variety of statistical analyses, one has to remember its startup message noting that it ``comes with ABSOLUTELY NO WARRANTY''.

Apart from the software issue, we provided a simple non-parametric example that an at first glance rather clever combination of some renown statistical methods (Theil's slopes and Tukeys's CI in our case) may yield disastrous results, if one ignores the assumptions of their usage.

%___________________
\section*{Appendix}

%___________
\noindent{\bf Proof of Theorem~\ref{theorem_theil_confidence}}\nopagebreak

Recall $N_c$, $N_d$ and the hypothesis $H_0$ from Section~\ref{section_theil}. In \citet{theil1950_part_I} on p. 390, the true confidence level of the Theil-type CI $(s_l,s_u)$ is expressed as
\[
1 - 2 \cdot P(N_d \leq l-1),
\]
where the probability is evaluated under $H_0$ and the result holds even under a more general setting than discussed in our paper. Since $K=N_c-N_d$ and $N_c+N_d=N$, we obtain that $N_d=(N-K)/2$ and
\[
1 - 2 \cdot P(N_d \leq l-1)
=
1 - 2 \cdot P\left( \frac{N-K}{2} \leq \left( \frac{N-k}{2}+1 \right)-1 \right)
=
1 - 2 \cdot P(K \geq k).
\]
\qed

%___________
\medskip
\noindent{\bf Proof of Theorem~\ref{theorem_o_vnoreni}}\nopagebreak

Since the smallest Walsh average $w_1$ is given by the smallest slope $s_1$ as $(s_1+s_1)/2=s_1$ and the largest Walsh average $w_{55}$ is given by the largest slope $s_{10}$ as $(s_{10}+s_{10})/2=s_{10}$, we obtain $s_1=w_1<w_9$ and $w_{47} < w_{55} = s_{10}$.\qed

%___________
\medskip
\noindent{\bf Proof of Theorem~\ref{theorem_o_t2_w9_w47_s_9}}\nopagebreak

{\it Part a):} We prove the equivalent statement ``$w_9<s_2$ iff $s_1+s_9 < 2s_2$''. Start with $w_9<s_2$ and consider the 9 smallest Walsh averages $w_1<w_2<\cdots<w_9$. Each of them is of the form $(s_i+s_j)/2$ for some $i \leq j$ and since $s_2=(s_2+s_2)/2$, the assumption $w_9<s_2$ means that
\begin{equation}\label{equivalent_statement}
 \frac{s_i+s_j}{2} < \frac{s_2+s_2}{2}.
\end{equation}
Because $s_1 < s_2 < \cdots < s_{10}$, the sharp inequality (\ref{equivalent_statement}) immediately implies, that $i=1$ and the 9 smallest Walsh averages $w_1<w_2<\cdots<w_9$ have to be of the form $(s_1+s_1)/2 < (s_1+s_2)/2 < \cdots < (s_1+s_9)/2$. Therefore, the inequality $w_9<s_2$ can be rewritten as $(s_1+s_9)/2 < (s_2+s_2)/2$ and the first part of the proof is complete.

Now, start with $s_1+s_9 < 2s_2$, i.e. $(s_1+s_9)/2 < s_2$. Since the 8 Walsh averages $(s_1+s_1)/2 < (s_1+s_2)/2 < \cdots < (s_1+s_8)/2$ are even smaller then $(s_1+s_9)/2$, we see that there are at least 9 Walsh averages smaller than $s_2$. Therefore, also the $9$-th smallest Walsh averages, i.e. $w_9$, is smaller than $s_2$.

{\it Part b):}  Note that the proof of part a) is based on the natural ordering ``the higher slope (or Walsh average), the higher index''. Using the reverse ordering ``the higher slope (or Walsh average), the lower index'' in the proof of part a), one obtains the ``symmetric'' counterpart of part a), which is part b).\qed

%___________
\medskip
\noindent{\bf Proof of Theorem~\ref{theorem_o_p1+p2+p3+p4}}\nopagebreak

Split the whole probability space into these four disjoint random events:
$$A: s_2\le w_9 \wedge w_{47}\le s_9$$
$$B: s_2\le w_9\wedge s_9< w_{47}$$
$$C: w_9<s_2\wedge w_{47}\le s_9$$
$$D: w_9<s_2 \wedge s_9<w_{47}$$
Denote by $U$ the random event $\{\beta_{1} \in (w_9,w_{47})\}$. Note that the minimum and the maximum of all slopes $s_i$ and their Walsh averages $w_i$ are $s_1$ and $s_{10}$, respectively. This implies that
\begin{eqnarray}
 P(U \cap A)
 & \leq &
 P( \{\beta_{1}\in (s_2,s_9)\} \cap A)
 =
 p_1,
\nonumber\\
 P(U \cap B)
 & \leq &
 P( \{\beta_{1}\in (s_2,s_{10})\} \cap B)
 =
 p_2,
\nonumber\\
 P(U \cap C)
 & \leq &
 P( \{\beta_{1}\in (s_1,s_9)\} \cap C)
 =
 p_3,
\nonumber\\
 P(U \cap D)
 & \leq &
 P( \{\beta_{1}\in (s_1,s_{10})\} \cap D)
 =
 p_4,
\nonumber
\end{eqnarray}
where the final equality in each row follows from Theorem~\ref{theorem_o_t2_w9_w47_s_9}. Hence, we obtain
\[
P(U)
=
(U \cap A) + P(U \cap B)+P(U \cap C)+P(U \cap D)
\leq
p_1+p_2+p_3+p_4.
\]
\qed

%___________
\medskip
\noindent{\bf Proof of Theorem~\ref{veta_obmedzujuca_usporiadania_sklonov}}\nopagebreak

By contradiction, let $S_{ac}$ be the greatest of $S_{ab}$, $S_{ac}$, $S_{bc}$ -- the case that $S_{ac}$ is the smallest can be treated analogously. By (\ref{S_pomocou_epsilonov}) and by noting that $x_a<x_b<x_c$, one observes that the inequality $S_{ac}>S_{ab}$ is equivalent to
\[
(\varepsilon_a-\varepsilon_c)(x_a-x_b)>(\varepsilon_a-\varepsilon_b)(x_a-x_c)
\]
and $S_{ac}>S_{bc}$ is equivalent to
\[
(\varepsilon_a-\varepsilon_c)(x_b-x_c)>(\varepsilon_b-\varepsilon_c)(x_a-x_c).
\]
By summing these two inequalities we obtain $(\varepsilon_a-\varepsilon_c)(x_a-x_c)>(\varepsilon_a-\varepsilon_c)(x_a-x_c)$ which is impossible.\qed

%___________
\medskip
\noindent{\bf Proof of Theorem~\ref{veta_nulovost_p3}}\nopagebreak

Theorem~\ref{veta_obmedzujuca_usporiadania_sklonov} implies that the minimum and maximum sample slopes $s_1$ and $s_{10}$ are of the form $s_1=S_{i,i+1}$ and $s_{10}=S_{j,j+1}$ for some distinct $i$ and $j$ from $\{1,2,3,4\}$. Straightforward algebra implies that $S_{i,j} - S_{i+1,j+1}=(s_{10} - s_1)/(i-j)$ under Condition~\ref{condition_ekvidistantnost}, which means that
\begin{equation}\label{s10-s1}
s_{10}-s_1
=
|i-j| \cdot |S_{i,j} - S_{i+1,j+1}|.
\end{equation}
Note that $|i-j| \leq 3$ (because $1 \leq i,j \leq 4$) and if both $S_{i,j}$ and $S_{i+1,j+1}$ belong to $\{s_2,s_3,\ldots,s_{9}\}$, then one obtains from (\ref{s10-s1}) that
\begin{equation}\label{s10-s1_zhora}
s_{10}-s_1
\leq
3(s_9-s_2).
\end{equation}
However, summing the inequalities $s_1+s_9 < 2s_2$ and $2s_9 < s_2+s_{10}$ appearing in the definition of $p_4$ yields
\[
s_{10}-s_1
>
3(s_9-s_2),
\]
which contradicts (\ref{s10-s1_zhora}), i.e. $p_4=0$.

It remains to treat the case when not both $S_{i,j}$ and $S_{i+1,j+1}$ belong to $\{s_2,s_3,\ldots,s_{9}\}$. This happens if and only if $|i-j|=1$. Without loss of generality, we will suppose that $j=i+1$, i.e. $s_1=S_{i,i+1}$, $s_{10}=S_{i+1,i+2}$ and $i\leq3$.

\emph{a) The case when $i \leq 2$.} We will show that the inequality
\begin{equation}\label{p3_nerovnost2}
2s_9 < s_2+s_{10}
\end{equation}
appearing in the definition of $p_4$ is impossible. Because $s_2 \leq S_{i+2,i+3}$, $S_{i+1,i+3} \leq s_9$ and $s_{10}=S_{i+1,i+2}$, the inequality (\ref{p3_nerovnost2}) would imply that $2 S_{i+1,i+3} < S_{i+2,i+3} + S_{i+1,i+2}$, which is equivalent to $0<0$ under Condition~\ref{condition_ekvidistantnost}.

\emph{b) The case when $i=3$.} We will show that the inequality
\begin{equation}\label{p3_nerovnost1}
s_1+s_9 < 2s_2
\end{equation}
appearing in the definition of $p_4$ is impossible. Because $s_2 \leq S_{2,4}$, $S_{2,3} \leq s_9$ and $s_1=S_{3,4}$, the inequality (\ref{p3_nerovnost1}) would imply that $S_{3,4} + S_{2,3} < 2 S_{2,4}$, which is equivalent to $0<0$ under Condition~\ref{condition_ekvidistantnost}.\qed

%___________
\medskip
\noindent{\bf Proof of Theorem~\ref{p2=p3}}\nopagebreak

Symmetry and independence of the distribution of the $\varepsilon_i$'s given by Condition~\ref{condition_rovnomerne_rozdelenie}, together with the equidistantness of the $x_i$'s given by Condition~\ref{condition_ekvidistantnost} means that moving from the $\varepsilon_i$'s to the ``equiprobable'' $-\varepsilon_i$'s reverts the ordering of the sample slopes and also the ordering of their Walsh averages, because each sample slope changes symmetrically around $\beta_1$ (cf. (\ref{S_pomocou_epsilonov})). It means that, for example, the sample slope with the label $s_2$ gets the label $s_9$, or the Walsh average with the label $w_9$ gets $w_{47}$, etc. The relationships between the $s_i$'s and $w_j$'s change accordingly: for example, $s_2 \leq w_9$ changes to $w_{47} \leq s_9$. Hence, we observe that the conditions defining $p_2$ change to conditions defining $p_3$.\qed

%___________________
\bibliographystyle{spbasic}
\bibliography{toth_somorcik_REVISION}

\begin{thebibliography}{28}
\providecommand{\natexlab}[1]{#1}
\providecommand{\url}[1]{{#1}}
\providecommand{\urlprefix}{URL }
\expandafter\ifx\csname urlstyle\endcsname\relax
  \providecommand{\doi}[1]{DOI~\discretionary{}{}{}#1}\else
  \providecommand{\doi}{DOI~\discretionary{}{}{}\begingroup
  \urlstyle{rm}\Url}\fi
\providecommand{\eprint}[2][]{\url{#2}}

\bibitem[{Arpaci et~al(2013)Arpaci, Eastaugh, and Vacik}]{*14*arpaci_etal2013}
Arpaci A, Eastaugh CS, Vacik H (2013) Selecting the best performing fire
  weather indices for \uppercase{A}ustrian ecoregions. Theoretical and Applied
  Climatology 114:393--406

\bibitem[{Barroso et~al(2015)Barroso, Nascimento, Nascimento, Fonseca~e Silva,
  Cruz, Bhering, and de~Paula~Ferreira}]{*18*barroso_etall2015}
Barroso LMA, Nascimento M, Nascimento ACC, Fonseca~e Silva F, Cruz CD, Bhering
  LL, de~Paula~Ferreira R (2015) Metodologia para an{\'a}lise de adaptabilidade
  e estabilidade por meio de regress{\~a}o quant{\'\i}lica. Pesquisa
  Agropecu{\'a}ria Brasileira 50:290--297

\bibitem[{B\"ueler and Enge(2003)}]{vinci}
B\"ueler B, Enge A (2003) Vinci.
  \urlprefix\url{http://www.math.u-bordeaux1.fr/$\sim$aenge/index.php?category=\linebreak
  software{\&}page=vinci}, version 1.0.5. Accessed 17 August 2016

\bibitem[{Carothers et~al(2010)Carothers, Goler, Kapoor, Lara, and
  Keasling}]{*23*carothers_etal2010}
Carothers JM, Goler JA, Kapoor Y, Lara L, Keasling JD (2010) Selecting
  \uppercase{RNA} aptamers for synthetic biology: investigating magnesium
  dependence and predicting binding affinity. Nucleic Acids Research
  38:2736--2747

\bibitem[{Cuevas et~al(2010)Cuevas, Calvo, Little, Pino, and
  Dassori}]{*13*cuevas_etal2010}
Cuevas JG, Calvo M, Little C, Pino M, Dassori P (2010) Are diurnal fluctuations
  in streamflow real? Journal of Hydrology and Hydromechanics 58:149--162

\bibitem[{Denys et~al(2012)Denys, Caboche, Tack, Rychen, Wragg, Cave,
  Jondreville, and Feidt}]{*24*denys_etal2012}
Denys S, Caboche J, Tack K, Rychen G, Wragg J, Cave M, Jondreville C, Feidt C
  (2012) In vivo validation of the unified barge method to assess the
  bioaccessibility of arsenic, antimony, cadmium, and lead in soils.
  Environmental Science \& Technology 46:6252--6260

\bibitem[{Eastaugh et~al(2012)Eastaugh, Arpaci, and
  Vacik}]{*05*eastaugh_etal2012}
Eastaugh CS, Arpaci A, Vacik H (2012) A cautionary note regarding comparisons
  of fire danger indices. Natural Hazards and Earth System Science 12:927--934

\bibitem[{Heiskanen et~al(2011)Heiskanen, Rautiainen, Korhonen, M{\~o}ttus, and
  Stenberg}]{*07*heiskanen_etal2011}
Heiskanen J, Rautiainen M, Korhonen L, M{\~o}ttus M, Stenberg P (2011)
  Retrieval of boreal forest \uppercase{LAI} using a forest reflectance model
  and empirical regressions. International Journal of Applied Earth Observation
  and Geoinformation 13:595--606

\bibitem[{Heiskanen et~al(2012)Heiskanen, Rautiainen, Stenberg, M{\~o}ttus,
  Vesanto, Korhonen, and Majasalmi}]{*10*heiskanen_etal2012}
Heiskanen J, Rautiainen M, Stenberg P, M{\~o}ttus M, Vesanto VH, Korhonen L,
  Majasalmi T (2012) Seasonal variation in \uppercase{MODIS LAI} for a boreal
  forest area in \uppercase{F}inland. Remote Sensing of Environment
  126:104--115

\bibitem[{Hollander and Wolfe(1999)}]{hollanderwolfe1999}
Hollander M, Wolfe DA (1999) Nonparametric Statistical Methods -- 2nd ed. John
  Wiley \& Sons, New York

\bibitem[{Hunter et~al(2012)Hunter, Veuger, and Witte}]{*21*hunter_etal2012}
Hunter WR, Veuger B, Witte U (2012) Macrofauna regulate heterotrophic bacterial
  carbon and nitrogen incorporation in low-oxygen sediments. The ISME journal
  6:2140--2151

\bibitem[{Hunter et~al(2013)Hunter, Jamieson, Huvenne, and
  Witte}]{*15*hunter_etal2013}
Hunter WR, Jamieson A, Huvenne VAI, Witte U (2013) Sediment community responses
  to marine vs. terrigenous organic matter in a submarine canyon.
  Biogeosciences 10:67--80

\bibitem[{Komsta(2013)}]{package_mblm}
Komsta L (2013) mblm: Median-Based Linear Models.
  \urlprefix\url{http://CRAN.R-project.org/package=\linebreak mblm},
  \uppercase{R} package version 0.12. Accessed 17 August 2016

\bibitem[{Kumari et~al(2012)Kumari, Nie, Chen, Ma, Stewart, Li, Lu, Taylor, and
  Wei}]{*06*kumari_etal2012}
Kumari S, Nie J, Chen HS, Ma H, Stewart R, Li X, Lu MZ, Taylor WM, Wei H (2012)
  Evaluation of gene association methods for coexpression network construction
  and biological knowledge discovery. PLoS One 7:e50,411

\bibitem[{Logan(2011)}]{*17*logan2011}
Logan M (2011) Biostatistical Design and Analysis Using R: A Practical Guide.
  Wiley--Blackwell, Chichester

\bibitem[{Lucas et~al(2013)Lucas, Sponseller, and Laudon}]{*19*lucas_etal2013}
Lucas RW, Sponseller RA, Laudon H (2013) Controls over base cation
  concentrations in stream and river waters: A long-term analysis on the role
  of deposition and climate. Ecosystems 16:707--721

\bibitem[{Mueller et~al(2014)Mueller, Dressler, Tucker, Pinzon, Leimgruber,
  Dubayah, Hurtt, B{\"o}hning-Gaese, and Fagan}]{*04*mueller_etal2014}
Mueller T, Dressler G, Tucker CJ, Pinzon JE, Leimgruber P, Dubayah RO, Hurtt
  GC, B{\"o}hning-Gaese K, Fagan WF (2014) Human land-use practices lead to
  global long-term increases in photosynthetic capacity. Remote Sensing
  6:5717--5731

\bibitem[{Pocewicz et~al(2007)Pocewicz, Vierling, Lentile, and
  Smith}]{*09*pocewicz_etal2007}
Pocewicz A, Vierling LA, Lentile LB, Smith R (2007) View angle effects on
  relationships between \uppercase{MISR} vegetation indices and leaf area index
  in a recently burned ponderosa pine forest. Remote Sensing of Environment
  107:322--333

\bibitem[{Puertas~Orozco et~al(2011)Puertas~Orozco, Carvajal~Escobar, and
  Quintero~Angel}]{*16*puertas_etal2011}
Puertas~Orozco OL, Carvajal~Escobar Y, Quintero~Angel M (2011) Study of monthly
  rainfall trends in the upper and middle cauca river basin,
  \uppercase{C}olombia. Dyna--Colombia 78:112--120

\bibitem[{{R Development Core Team}(2010)}]{eRko}
{R Development Core Team} (2010) R: A Language and Environment for Statistical
  Computing. R Foundation for Statistical Computing, Vienna, Austria,
  \urlprefix\url{http://www.R-project.org}, {ISBN} 3-900051-07-0

\bibitem[{Sardans and Pe{\~n}uelas(2015)}]{*11*sardans_penuelas2015}
Sardans J, Pe{\~n}uelas J (2015) Trees increase their
  \uppercase{P}:\uppercase{N} ratio with size. Global ecology and biogeography
  24:147--156

\bibitem[{Smith(1967)}]{smith1967}
Smith EJ (1967) Cloud seeding experiments in \uppercase{A}ustralia. In:
  Proceedings of the Fifth Berkeley Symposium on Mathematical Statistics and
  Probability, University of California Press, Berkeley, California, USA, vol
  5: Weather Modification, pp 161--176

\bibitem[{Theil(1950{\natexlab{a}})}]{theil1950_part_I}
Theil H (1950{\natexlab{a}}) A rank-invariant method of linear and polynomial
  regression analysis, \mbox{I}. Proceedings of the Koninklijke Nederlandse
  Akademie van Wetenschappen 53:386--392

\bibitem[{Theil(1950{\natexlab{b}})}]{theil1950_part_II}
Theil H (1950{\natexlab{b}}) A rank-invariant method of linear and polynomial
  regression analysis, \mbox{II}. Proceedings of the Koninklijke Nederlandse
  Akademie van Wetenschappen 53:521--525

\bibitem[{Theil(1950{\natexlab{c}})}]{theil1950_part_III}
Theil H (1950{\natexlab{c}}) A rank-invariant method of linear and polynomial
  regression analysis, \mbox{III}. Proceedings of the Koninklijke Nederlandse
  Akademie van Wetenschappen 53:1397--1412

\bibitem[{Vannest et~al(2013)Vannest, Davis, and Parker}]{*08*vannest_etal2013}
Vannest KJ, Davis JL, Parker RI (2013) Single Case Research in Schools:
  Practical Guidelines for School-Based Professionals. Taylor \& Francis, New
  York

\bibitem[{Wheeler(2009)}]{package_Supp_Dists}
Wheeler B (2009) SuppDists: Supplementary distributions.
  \urlprefix\url{http://CRAN.R-project.org/package=\linebreak SuppDists},
  \uppercase{R} package version 1.1-8. Accessed 17 August 2016

\bibitem[{Zottele et~al(2010)Zottele, Toller, and Eccel}]{*03*zottele_etal2010}
Zottele F, Toller G, Eccel E (2010) Irri4web: crop water needs definition by
  web\uppercase{GIS}. Italian Journal of Agrometeorology 14:5--14

\end{thebibliography}

\end{document}